\documentclass[aps,prb,twocolumn,showpacs,superscriptaddress,preprintnumbers,amsmath,amssymb]{revtex4}

\usepackage{txfonts}
\usepackage{amssymb}

\usepackage{graphicx}
\usepackage{dcolumn}
\usepackage{bm}
\usepackage{color}

\begin{document}


\title{Valley detection using a graphene gradual pn junction with spin-orbit coupling: an analytical conductance calculation}

\author{Mou Yang}
\altaffiliation{Electronic address: yang.mou@hotmail.com}
\affiliation{Guangdong Provincial Key Laboratory of Quantum Engineering and Quantum Materials, \\
School of Physics and Telecommunication Engineering,
South China Normal University, Guangzhou 510006, China}

\author{Rui-Qiang Wang }
\affiliation{Guangdong Provincial Key Laboratory of Quantum Engineering and Quantum Materials, \\
School of Physics and Telecommunication Engineering,
South China Normal University, Guangzhou 510006, China}

\author{Yan-Kui Bai}
\affiliation{College of Physical Science and Information Engineering and Hebei Advance Thin Films Laboratory, Hebei Normal university, Shijiazhuang, Hebei 050024, China}

\begin{abstract}
Graphene pn junction is the brick to build up variety of graphene nano-structures. The analytical formula of the conductance of graphene gradual pn junctions in the whole bipolar region has been absent up to now. In this paper, we analytically calculated that pn conductance with the spin-orbit coupling and stagger potential taken into account. Our analytical expression indicates that the energy gap causes the conductance to drop a constant value with respect to that without gap in a certain parameter region, and manifests that the curve of the conductance versus the stagger potential consists of two Gaussian peaks -- one valley contributes one peak. The latter feature allows one to detect the valley polarization without using double-interface resonant devices.
\end{abstract}


\maketitle

\section{Introduction}

Since graphene was discovered, many interesting physical effects were predicted and some were experimentally confirmed.\cite{graphene,review}
Graphene nowadays is regarded as the new-generation material and draws extensive attentions. The pn junction is the simplest structure in graphene system and is the brick to build up more complicated structures such as pn superlattice,\cite{superlattice} quantum dot,\cite{dot} and point contact.\cite{contact} Lots of intriguing effects are induced by the pn junction, for examples, Klein tunneling for normally injected electrons,\cite{review} negative refraction and Veselago lens for electron beams,\cite{lens,lens-supper}, snake-Hall states around the interface under strong magnetic field,\cite{snake,snake1}, and non-integer conductance quantization due to the mix of electron and hole Landau edge modes near the pn interface.\cite{mix,mix1} 

In Ref. [\onlinecite{pn-junction}], an analytical formula for the graphene pn conductance was proposed for the special case that the Fermi energy lies at the middle point between the potentials at two ends. However, the analytical calculation of pn junctions for the whole bipolar region has been lacked for a long time. Furthermore, spin-orbit coupling (SOC) was not included in the calculation of Ref. [\onlinecite{pn-junction}], which plays more and more important role in nowadays physics society. It is well known that the SOC is too weak to be observed in pristine graphene.\cite{review} However, the SOC deserves to be considered because it can be dramatically magnified to measurable scale by a few methods.\cite{hydrogenate} Moreover, a family of graphene-like materials such as silicene, germanene, and stanene (single layer of silicon, germanium, and stannum) were recently reported to have the same honeycomb lattice as graphene but two types of sub-lattices buckled up and down out of plane. The lattice buckling leads to observable SOC,\cite{evidence,gap-SOC,gap-tune,on-graphene} which make these graphene and graphene-like materials be topological insulators.\cite{Kane} By including the SOC and stagger potential, the graphene lattice model is not a problem for graphene itself, but a platform to study different types of junctions\cite{domain,junction,spin-valley} and varies phases of two-dimensional topologic insulators.\cite{polarization,multiple-phases,interface,anomalous}

\begin{figure}
\includegraphics[width=7cm]{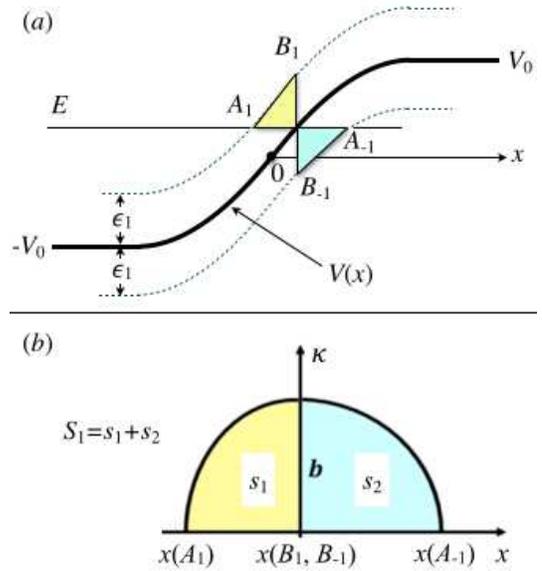}
\caption{(Color Online) \label{area}
(a) Sketch of the tunneling when electrons pass the pn junction. (b) The area of enclosed by $\kappa$ in the $x$-$\kappa$ plane in the process of an electron transmiting through the pn junction.}
\end{figure}

Motivated by these reasons, we studied the conductance of gradual graphene pn junctions with the SOC considered, which induces a energy gap and drives graphene to be a topologic insulator. By means of the subband tunneling model\cite{Yang}, we derived an analytical conductance expression of the graphene pn junction in the whole interval of the bipolar region. To check the validity of our analytical result, we numerically calculated the pn conductance using the non-equilibrium Green's function (NEGF) method, and found the conductances obtained by the two methods fit each other strikingly well. According to the analytical result, there exists a parameter interval, in which the conductance drops a constant value with respect to that without gap. When the stagger potential is present, the energy gaps for valleys $K$ and $K'$ become different. The analytical expression allows us to decomposes the conductance into valley $K$ and $K'$ components. The conductance is always governed by the valley having the smaller gap. The conductance versus the stagger potential consists of two Gaussian peaks, which are contributed by the two valleys. This effect can be used for valley detection in the graphene-like materials without using double-interface resonant devices.

\section{Subband Transmission}

The dispersion of graphene or graphene-like lattice around one valley with a uniform background potential $V$ reads 
\begin{eqnarray}\label{bulk}
\hbar^2v_F^2(k_x^2 + k_y^2)+\Delta^2 =(E - V)^{2},
\end{eqnarray}
where $\hbar$ is the reduced Plank constant, $v_F$ is the Fermi velocity, and $\boldsymbol{k}=(k_x,k_y)$, $2\Delta$, $E$ and $V$ are the wave-vector, energy gap, energy of electron, and background potential. Here we has not specified the physical origin of the energy gap, it can be induced by stagger potential, SOC, or the combination of both. For the electrons in a ribbon along $x$-direction, $x$-component wave-vector is a good quantum number, and the motion in $y$-direction is described by lateral mode index $n$. The dispersion relation of a graphene ribbon is
\begin{eqnarray}\label{ribbon}
\epsilon_{n}^{2} + \hbar^2v_F^2k_{n}^{2}=(E - V)^{2},
\end{eqnarray}
where $\epsilon_n$ is the lateral energy of mode $n$, which is positive for electrons and negative for holes, and $k_n$ is longitudinal wave-vector of electron of subband $n$. In the equation, $\Delta$ does not appear because its effect is absorbed in $\epsilon_n$. The minimum or maximum energies of the electron or hole dispersions for a given mode $n$ is regarded as the subband edge, which is obtained by letting $k_n=0$ as
\begin{eqnarray}\label{edge}
E_{n}&=&\epsilon_{n}+V.
\end{eqnarray}
When the ribbon is subjected by a pn junction potential $V(x)$, the uniform potential $V$ in Eqs. (\ref{ribbon}) and (\ref{edge}) is replaced by $V(x)$, and the subband edges are modulated by the pn potential, as illustrated in Fig \ref{area} (a). In this case $k_n$ is no longer a good quantum number and varies when electron propagating, saying, $k_n=k_n(x)$.

Now we track the physics when the electron goes through the pn junction with the help of Fig. \ref{area} (a). We consider an electron-like particle of lateral mode $n$ ($n=1$ in the figure) starts from $x=-\infty$. Before it is injected in the pn region, it has a real wave-vector $k_n$. After it enters the pn area, the subband edge $E_n$ increases, $k_n$ decrease and goes into zero at the turning point $A_n$. After it moves beyond the point, $k_n$ becomes imaginary (the electron goes evanescently with the evanescent wave-vector $\kappa_n=|k_n|$), and $k_n^2$ becomes more and more negative. This indicates $(E-V)^2$ in Eq. (\ref{ribbon}) decreases continuously, and runs into zero at point $B_n$. The electron can no longer travel on of mode $n$ and has to transit to subband $-n$ as a hole, i.e., from point $B_n$ to $B_{-n}$. 
The evanescent wave propagation continues until the electron 
reaches the turning point $A_{-n}$ on the curve of subband edge $-n$, the wave-vector recovers to a real value, and propagates away. According to the WKB theory, the transmission is determined by the evanescent transport part as 
\begin{eqnarray}\label{WKB}
T_{n} = e^{-2S_n}, 
\end{eqnarray}
where $S_n$ stands for the area enclosed by $\kappa_n(x)$ and $\kappa_{-n}(x)$ in the $x$-$\kappa$ plane.
We can approximate the curves of $\kappa_n$ and $\kappa_{-n}$ on the $x$-$\kappa$ plane as two quarters 
of two different ellipses \cite{addition} sharing one principle radius $b=|\epsilon_n|$. The other radius of each quarter is determined by the explicit form of $V(x)$. The two ellipse quarters are shown in Fig. \ref{area} (b). 
Letting $d_n$ be the distance between $A_n$ and $A_{-n}$ in $x$-direction, the area of the two quarters reads  
\begin{eqnarray}\label{S}
S_n &=& \frac \pi 4  \frac{d_n |\epsilon_n|}{\hbar v_F},
\end{eqnarray}
The conductance of the pn junction is thus the summation of the contributions of all the subband involved into the pn tunneling. 

\section{Analytical calculation of conduction}

\begin{figure}
\includegraphics[width=8cm]{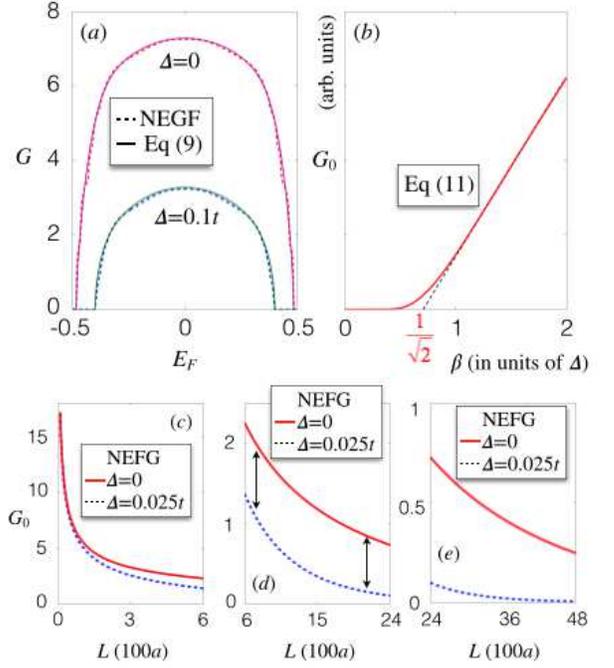}
\caption{(Color Online) \label{betaL}
(a) Conductance $G$ (in units of $e^2/h$) for $L=30a$ versus $E_F$, where $a$ is the in-plane atom-atom distance. (b) $G_0$ ($G$ at $E_F=0$) in Eq. (\ref{G0}) versus $\beta$. (c)-(e) Numerical calculated $G_0$ (in units of $e^2/h$) versus $L$ (in units of $100a$). Parameters are $W=100\sqrt3 a$ and $V_0=0.5t$ with $t$ the nearest hopping energy in tight-binding model. }
\end{figure}
The theory of the transmission for a pn junction of a ribbon can be applied to the bulk system. Given an electron injected with the wave-vector $\boldsymbol{k}$, the lateral energy $\epsilon_n$ should be replaced with $\pm(\hbar^2 v_F^2 k_y^2 + \Delta^2)^{1/2}$, where plus for electrons and minus for holes. The distance between the turning points $A_n$ and $A_{-n}$ (see Fig. \ref{area}) depends on not only the lateral energy but the potential slope at the incident energy
\begin{eqnarray}\label{F}
F = \left| \frac{dV}{dx} \right|.
\end{eqnarray}
After substituting $d_n$ by $d\approx2(\hbar^2 v_F^2k_y^2 + \Delta^2)^{1/2}/F$, we have the transmission
\begin{eqnarray}\label{T}
T \approx \exp\left(-\frac{\Delta^2}{\beta^2}
\right) \exp\left[-\frac{(\hbar v_F k_y)^2}{\beta^2}
\right],
\end{eqnarray}
where $\beta$ is a parameter of dimension of energy
\begin{eqnarray}\label{beta}
\beta = \left(\frac{\hbar v_F}{\pi}\right)^{\frac12} \cdot \sqrt{F}.
\end{eqnarray}
The left and right end (n-end and p-end) are assumed to be lifted by $-V_0$ and $V_0$ respectively. 
If we set $\Delta=0$ and $E=0$, by defining the effective length $l=V_0/F=\hbar v_F k/F$ and the incident angle $ \theta=\arctan (k_y/k_x)$, we recover the formula appearing in Ref. [\onlinecite{pn-junction}], $T=e^{-\pi kl \sin^2\theta}$. 

By counting all the transmission of different $k_y$, we have the conductance across the pn juntion,
\begin{eqnarray}\label{G}
G &=& \frac{e^2}{h}\int_{k_F} T \frac{W}{2\pi}dk_y = \frac{e^2}{h}\frac{W}{2\sqrt\pi} \times \nonumber \\
&& {\beta}  \exp\left(-\frac{\Delta^2}{\beta^2}
\right) \text{erf}\left(\frac{\sqrt{(V_0-|E_F|)^2-\Delta^2}}{\beta}\right),
\end{eqnarray}
where $W$ is the width of the sample and erf represents the error function. In the equation, it has to bear in mind that $\beta$ is a function of energy, saying, $\beta = \beta(E)$. 

To examine the validation of Eq. (\ref{G}), we numerically calculated conductance of graphene pn junctions for comparison.  The pn potential is modeled by a half period of sinusoidal function of the length $L$ connecting the two ends.
The numerical conductance is obtained by the NEGF method based on the tight-bonding Hamiltonian including the Kane-Mele SOC. The SOC induces an energy gap $\Delta=\Delta_{SO}$, where $\Delta_{SO}$ is a parameter to characterize the SOC amplitude in the tight-binding Hamiltonian. To eliminate the boundary effect and simulate the infinite system, the periodical edge condition is adopted. The details of numerical NEGF calculation and the tight-binding Hamiltonian are not shown here but presented in the Appendix for completeness. Figure \ref{betaL} (a) shows the conductances calculated from numerical and analytical methods.
As one can see, they are consistent very well. These conductance curves are peaked and centered at $E_F=0$ with peak width $2(V_0-\Delta)$. Larger energy gap means larger area of ellipses shown in Fig. \ref{area} (b), and leads to smaller conductance. Eq. (\ref{G}) is obtained from the WKB approximation, so the consistency will break down for the potential slope larger than a criterion value $F_c$. By comparison the conductance for many choices of $L$, 
we estimate the criterion slope to be $F_c\approx 0.25t/a$, where $t$ is 
the nearest hopping energy and $a$ is the in-plane atom-atom distance. For the graphene lattice, $t=2.7$eV, $a=1.4$\AA, so we have $F_c=5\times10^3$V/$\mu$m, and similar values of $F_c$ for other graphene-like materials, much greater than the experimentally achieved value, which is only several V/$\mu$m.\cite{fabricate,fabricate1} In realistic situations, the slope of pn potentials always allow the WKB approximation to be applied safely.

Particularly, if the Fermi energy is aligned at the middle energy between the potentials of two ends, saying, $E_F=0$, the conductance reaches its maximum 
\begin{eqnarray}\label{G0original}
G_0 = \beta \exp\left(-\frac{\Delta^2}{\beta^2}
\right) \text{erf}\left(\frac{\sqrt{V_0^2-\Delta^2}}{\beta}\right). 
\end{eqnarray} 
In the equation and from here on, we omit the constant pre-factor $e^2W/(2\sqrt\pi h)$ for simplicity. Because the potential varies slowly along $x$, we consider the situation $\beta \ll  (V_0^2-\Delta^2)^{1/2}$. For this case the error function gives 1, and the conductance is reduced to
\begin{eqnarray}\label{G0}
G_0 =  \beta \exp\left(-\frac{\Delta^2}{\beta^2}\right).
\end{eqnarray} 
When no bulk gap exists, saying, $\Delta=0$, it is interesting to note the conductance is simply proportional to $\beta$. Figure \ref{betaL} (b) shows the conductance at $E_F=0$ as a function of the junction length. The conductance remains almost non-changed when $\beta$ below a threshold, and then changes linearly with $\beta$ increases. In this special linear region, the conductance reads
\begin{eqnarray}\label{G0para}
G_0 \sim \left(\beta-\frac{\Delta}
{\sqrt2}\right), \quad \frac{\Delta}{\sqrt2}<\beta<\sqrt2 \Delta.
\end{eqnarray} 
When $\beta$ is out of the region, say, when much smaller than the lower limit, no conductance can be observed, and when much larger than the upper limit, the exponential term in Eq. (\ref{G0}) tends to unity and the conductance is proportional to $\beta$. Equation (\ref{G0para}) means that the conductance of a system with an energy gap is smaller than that of gapless system by a constant value in the parameter region specified in the equation. This interval of $\beta$ can be translated to a scope of the junction length as,
\begin{eqnarray}\label{L}
\frac12 L_c < L < 2 L_c,
\end{eqnarray} 
where $L_c=V_0/\Delta^2$. Figures \ref{betaL} (c) through (e) show numerically calculated $G_0$ versus $L$ for both the cases $\Delta=0$ and $\Delta \neq 0$. For the case of $V_0=0.5t$ and $\Delta=0.025t$, we have $L_c=1200a$. In the interval of $L$ from $600a$ through $2400a$, the difference between the two conductances is almost a constant, as illustrated in Fig. \ref{betaL} (d).

\section{Valley featured transport}

\begin{figure}
\includegraphics[width=8cm]{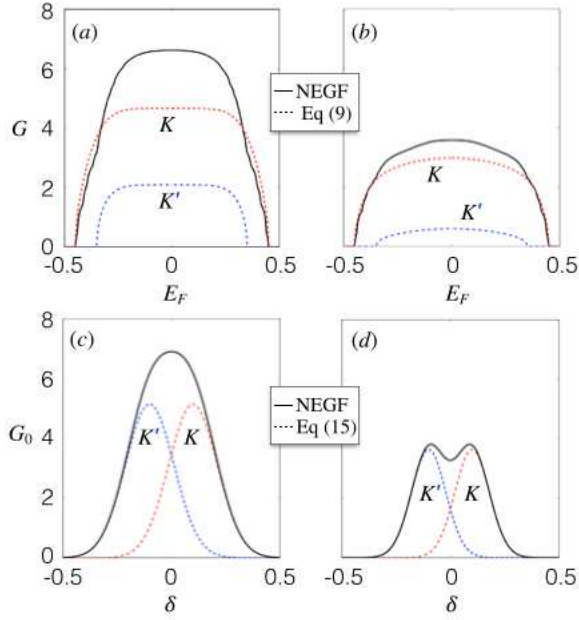}
\caption{(Color Online) \label{GED}
Conductance and its valley components (in units of $e^2/h$) as functions of $E_F$ (in units of $t$) for (a) $L=30a$ and (b) $L=60a$. Conductance and its valley components (in units of $e^2/h$) at $E_F=0$ as functions of $\delta$ (in units of $t$) for (c) $L=30a$ and (d) $L=60a$. The curves of total conductance are calculated numerically and the conductance components of valleys are calculated analytically. Other parameters are $W=100\sqrt3a$ and $V_0=0.5t$. }
\end{figure}

If there is a stagger potential present, saying, potentials on A-type and B-type atoms are lifted by $\delta$ and $-\delta$ respectively. The stagger potential leads to the electron dispersion and the hole dispersion come closer for one valley and leave apart for the other valley. Thus, the energy gaps of the two valleys are driven to be different, say,
\begin{eqnarray}\label{DeltaK}
\Delta_{K,K'}= |\delta \mp \Delta_{SO}|.
\end{eqnarray}
The valley dependent energy gap results in that the conductances at $E_F=0$ for different valley are 
\begin{eqnarray}\label{GK}
G_0^{K,K'} = \beta \exp \left[-\frac{(\delta\mp \Delta_{SO})^2}{\beta^2}\right].
\end{eqnarray}
The total conductance is $G_0 = G_0^{K} + G_0^{K'}$. This equation reveals that when we treat the stagger potential as a tunable parameter, the conductance component of a single valley as a function of the stagger potential is a Gaussian peak, which is centered at 
$\delta=\Delta_{SO}$ or $-\Delta_{SO}$, depending on the valley, and the full width at half maximum (half-width) of which is $2\beta \sqrt{\ln2}$. 

The peak phenomenon of the conductance has two important features. First, the maximum conductance is proportional to $\beta$, and for a given pn potential height $V_0$, it is proportional to $1/\sqrt L$. So the peak will survive for super long pn junctions and one need not to worry about the conductance signal fades out. Second, the peak feature is not a resonant effect, and need not tune the parameters to meet the resonant condition.

Figure \ref{GED} (a) through (d) shows the conductance and the conductance components of valleys $K$ and $K'$ of pn junctions as functions of the Fermi energy and as functions of of stagger potential amplitude $\delta$. The conductance contributed by both valleys is calculated numerically, and the conductance component of each valley is analytically. In Fig. \ref{GED} (a), it can be clearly seen that the conductance consists of two peaks with different peak widths. The two peaks are contributed by valleys $K$ and $K'$, and the valley with the smaller gap leads to the wider and higher peak. For a longer pn junction, the decrease of the conductance component of valley $K'$ is more notably than that of valley $K$, as shown in Fig. \ref{GED} (b). The conductance is always dominated by the transport of the valley with the smaller gap. In Fig. \ref{GED} (c), the conductance curve shows a peak with maximum at $\delta=0$, but it indeed consists of two peaks of Gaussian type. For the parameters in Fig. \ref{GED} (c), the half-width is about $0.26t$, greater than the peak separation $2\Delta_{SO}=0.2t$, and the two peaks cannot be distinguished on the conductance curve. Figure \ref{GED} (d) shows the conductance and components for a longer pn junction. Since longer $L$ means smaller $\beta$ and thus narrower peaks (half-width is about $0.186t$, smaller than the peak separation $2\Delta_{SO}$), the resolved peaks appear on the total conductance curve. One can choose a longer $L$ so as to the two peaks are totally resolved. All the above discussions are based on single spin. Indeed The conductance for the other spin is the same, i.e., $G^{\uparrow}=G^{\downarrow}$. Because flipping the spin leads to the interchange of the gaps of valleys $K$ and $K'$, the conductance component of valley $K$ for spin-up and that of valley $K'$ for spin-down are equivalent, i.e., $G^{K,\uparrow}=G^{K',\downarrow}$. 

The peak-behavior of the conductance can be used to detect valley index of electric current and to measure the valley-polarization by combining a pn junction with a spin-filter. Freely tuning the stagger potential may be a challenge for graphene, but is easy for other graphene-like materials such as silicene, germanene and stanene, due to the lattice buckling. Many schemes of valley detection proposals in other literatures rely on valley-dependent resonance of double-interface scattering, which needs accurate parameter control and is difficult to result in reliable observation. 

\section{Summary}

We obtained an analytical expression of the conductance of graphene pn junctions in the bipolar region with the SOC included and verified its validity by comparing it and the numerically calculated conductance. Our analytical result indicates that the conductance as a function of stagger potential consists of two Gaussian peaks,
which are contributed by the tunneling of electrons in two valleys respectively. This effect can be used for valley detection.

\acknowledgements

This work was supported by NSF of China Grant No. 11274124, No. 11474106, No. 11174088, and Hebei NSF Grant No. A2012205062.

\appendix
\section{Tight-binding Hamiltonian and numerical calculation of conductance}
The tight-binding Hamiltonian of a perfect graphene ribbon including the SOC reads
\begin{eqnarray}\label{H0}
H_0 &=& - t\sum_{\langle ij\rangle \alpha} c_{i\alpha}^+ c_{j\alpha}+ i\gamma\sum_{\langle\langle ij\rangle\rangle \alpha\beta} \nu_{ij} c_{i\alpha}^+ \sigma_{\alpha\beta}^z c_{j\beta}, 
\end{eqnarray}
where $c_{i\alpha}^+$ ($c_{i\alpha}$) is the creation (annihilation) operator for an electron with spin $\alpha$ on site $i$, $\sigma^z$ is the $z$-component of Pauli matrix, and the summations with the brackets $\langle .. \rangle$ and $\langle \langle .. \rangle \rangle$ run over all the nearest and next-nearest neighbor sites, respectively. The first term is the Hamiltonian of the nearest neighbor hopping with hopping energy $t$. The second term is the SOC Hamiltonian which involves the next-nearest neighbor hopping with amplitude $\gamma$ and a path dependent factor $\nu_{ij}$. For the electron couples form atom $i$, mediated by a nearest neighbor site and to a next-nearest neighbor atom $j$, we have $\nu_{ij}=1$ if it makes a left turn and $\nu_{ij}=-1$ if goes a right turn. 
The SOC term results in an energy gap as 
\begin{eqnarray}\label{DeltaSO}
\Delta_{SO} = 3\sqrt3 \gamma.
\end{eqnarray}
The Hamiltonian including the stagger potential, if it is present,   reads
\begin{eqnarray}\label{H0}
H_1 &=& H_0+\delta \sum_{i\alpha} \mu_{i}c_{i\alpha}^+ c_{i\alpha},  
\end{eqnarray}
where $\delta$ is the stagger potential amplitude, and $\mu_i=1$ when $i$ represents an A-type atom and $\mu_i=-1$ for a B-type atom. The stagger potential itself means a gap $|\delta|$. The combination of the stagger gap and the SOC gap results in two different gaps for valleys $K$ and $K'$, as in Eq. (\ref{DeltaK}).
When the ribbon is subjected by a space-varying potential, the Hamiltonian is 
\begin{eqnarray}\label{H}
H = H_1+\sum_{i\alpha} V_{i}c_{i\alpha}^+ c_{i\alpha}, 
\end{eqnarray}
where $V_i$ is the space-varying potential on site $i$.  
The conductance can be calculated by the NEGF method as 
\begin{eqnarray}\label{NEGF}
G={\rm Tr}(\Gamma_L G^r \Gamma_R G^a),
\end{eqnarray} 
where $G^{r(a)}$ and $\Gamma^{L(R)}$ are the retarded (advanced) Green's function of the area covered by the pn potential and the linewidth function of the left (right) lead. All these functions are obtained through the tight-binding Hamiltonian.


\begin{references}
\bibitem{addition} The area $s_1$ in Fig. \ref{area} (b) can be approximated as an ellipse quarter is because the derivatives $d\kappa/dx$ at points $A_1$ and $B_1$ are $\infty$ and $0$ respectively.
\end{references}
\end{document}